# Nitrogen ion beam synthesis of InN in InP(100) at elevated temperature

S. Dhara,[a] P. Magudapathy, R. Kesavamoorthy, S. Kalavathi, V. S. Sastry and K. G. M. Nair

Materials Science Division, Indira Gandhi Centre for Atomic Research, Kalpakkam-603 102

G. M. Hsu, L. C. Chen and K. H. Chen[b]

Centre for Condensed Matter Sciences, National Taiwan University, Taipei-106, Taiwan

K. Santhakumar and T. Soga

Department of Environmental Technology and Urban Planning, Nagoya Institute of

Technology, Nagoya 466-8555, Japan

## Abstract

InN phase is grown in crystalline InP(100) substrates by 50 keV $N^+$ implantation at an elevated temperature of 400 $^0$C followed by annealing at 525 $^0$C in $N_2$ ambient. Crystallographic structural and Raman scattering studies are performed for the characterization of grown phases. Temperature- and power-dependent photoluminescence studies show direct band-to-band transition peak ~1.06 eV at temperatures ≤150K. Implantations at an elevated temperature with a low ion beam current and subsequent low temperature annealing step are found responsible for the growth of high-quality InN phase.

[a] Corresponding author Email : dhara@igcar.gov.in

b) Also affiliated to Institute of Atomic and Molecular Sciences, Academia Sinica, Taipei 106, Taiwan

III-V nitrides have, recently, attracted great attention and are intensively studied as materials for bright light emitters that cover continuously from the ultraviolet to near-infrared region by proper alloying. With newly found direct bandgap of ~0.7 eV in InN, possibility of bandgap tuning in InGaN covers the range of 0.7-3.4 eV.[1] Large drift velocity in InN at room temperature makes it a better choice of material than GaAs and GaN for field effect transistors.[2] InN/Si tandem cells have been proposed for high-efficiency solar cells.[3] Nitridation of the In is generally difficult since the sticking probability of nitrogen on the surface of In is very small and the formed InN phase disintegrate above 600 $^0$C.[4] There are several reports where nitridation by equilibrium and non-equilibrium techniques have been employed to prepare the InN surfaces. In these methods, nitrogen source is used to react with In on sapphire substrate to form InN at low temperature to keep the nitrogen content at or close to stoichiometry.[4,5] Formation of embedded structure of III-V nitrides by ion beam implantation technique is an emerging field. Numerous studies are being carried out to synthesis the embedded structure by ion beam technique, as it is not feasible by conventional methods. Formation of GaN is reported in GaAs and GaP substrates in a two-step technique exploiting N$^+$ implantation at elevated temperatures and subsequent annealing around 900$^0$C.[6,7,8] A few nitridation work on InP is carried out by implantation at room temperature, where a similar two-step technique is adopted with the formation of an intermediate InP$_x$N$_{1-x}$ phase and subsequent annealing in N$_2$ ambient to grow a InN surface layer.[9,10] Except for the formation of chemical bonds, detailed structural studies are missing in these studies. A single-step technique is reported by our group for the formation of InN phase by implanting high N$^+$ beam current (~25.5 μA cm$^{-2}$) in InP substrates at an elevated temperature of 400 $^0$C.[11] In this technique a high N$^+$ beam current is used to disintegrate In-P bonds and simultaneous formation of In-N bonds in the dynamic annealing (defect annihilation) process.

However, dynamic annealing process is never perfect and the Raman scattering study showed poor crystalline quality of the grown InN phase.

In the present study, our main motivation is to synthesize high-quality InN layer and to study the evolution of its growth from InP phase. A similar technique of $N^+$ implantation in InP substrate at an elevated temperature, as adopted for the single-step process in our previous study,[11] is used for the growth InN layer. The major difference from the previous study is the use of a low ion beam current to avoid any dynamic annealing during the implantation process. The sequence of transformation of InP to InN is an important aspect to be studied to probe the phase evolution and its dependence on growth orientation.

50 keV $N^+$ implantation in crystalline InP(100) substrates at 400 $^0$C is performed using a low energy ion implanter with gaseous ion source. Implantation at elevated temperature is planned to minimize defects introduced during energetic implantation process. A low beam current of ~ 1.8 $\mu$A cm$^{-2}$ is used to achieve fluences in the range of $1 \times 10^{17}$ – $4 \times 10^{17}$ cm$^{-2}$. As calculated from the SRIM code,[12] the mean projected range of 50-keV $N^+$ in InP is ~100 nm and longitudinal straggling is ~70 nm. So, grown layer is having a thickness around 70 nm distributed over a depth of 65 -135 nm from the surface of InP substrate. Implanted samples are annealed at low temperature of 525 $^0$C for 15 minutes in nitrogen ambient, in order to avoid disintegration of In-N bond (reported to disintegrate above 600 $^0$C).[4] The evolution of phases created by ion bombardment is studied using the grazing incidence X-ray diffraction (GIXRD; STOE Diffractometer) technique. All the measurements are carried out with an angle of incidence $\theta_{inc}$= 0.3$^o$ for X-rays. Raman spectra of the as-grown and implanted samples are recorded in back scattering geometry using 514.5 nm Ar$^+$ laser line in the range of 470 - 570 cm$^{-1}$ for probing modes corresponding to the InN phase.

Temperature- and power-dependent photoluminescence (PL) studies using $Ar^+$ laser with excitation wavelength of 488 nm are performed to measure, as well as to confirm the direct band-to-band peak position in the grown phase.

The GIXRD patterns of the unimplanted and post-implanted annealed samples are shown in Fig. 1. The peaks at $2\theta$ values of $30.43^0$ and $63.36^0$ correspond to reflections from (200) and (400) planes of unimplanted crystalline InP substrates. The GIXRD pattern for the annealed sample implanted with a fluence of $2x10^{17}$ $cm^{-2}$ shows peaks at $2\theta$ values of $27.26^0$, $52.16^0$, and $56.09^0$ (marked as dot in the figure) correspond to $InP_xN_y$ phase, as concluded from the peak positions falling in between locations corresponding to reflections from InN and InP phases.[13] This may be due to the partial nitridation of In-P bonds at low fluences. The peak at $51.6^0$ corresponds either to (311) plane of InP or (110) of InN phases (Fig. 1).[13] Most interesting observation is that the same peak is intensified by more than an order of magnitude for the annealed sample implanted at a high fluence of $4x10^{17}$ $cm^{-2}$. The contribution from InP (311) plane can not be ruled out for the diffraction peak at $51.6^0$, particularly for the sample grown in the non-equilibrium process of ion beam technique. We may also restate that the grown InN phase is a sandwiched layer inside the InP substrate, so observing diffraction signal corresponding to InP phase from that of the InN phase is unavoidable even by GIXRD technique. Nitridation of the sample by ion implantation and annealing treatment in the $N_2$ ambient, however, point to the possibility of major contribution from the (110) plane of InN phase. Observation of only one peak corresponding to InN phase in the sample implanted at a high fluence also indicates the possibility of texturing in the grown layer. Increase in the diffracted intensity by more than an order in the sample grown at high fluence is also supportive to the assumption of texturing.

Raman scattering study of the annealed sample irradiated at a fluence of $2 \times 10^{17}$ cm$^{-2}$ shows (Fig. 2a) peaks around 490 cm$^{-1}$, 505 cm$^{-1}$ and 530 cm$^{-1}$, which correspond to transverse optical mode of $E_1$(TO), $E_2$(high) and $B_1$(high), respectively, of the wurtzite InN phase.[10] The $E_1$(TO) mode is reported in the range of 475-490 cm$^{-1}$ depending on the crystalline quality of InN with the theoretical prediction of 470 cm$^{-1}$ [ref. 14] for ideally good quality crystal. Peaks around 515 cm$^{-1}$ and 540 cm$^{-1}$ correspond to disorder assisted longitudinal optical (DALO) modes arising from the InN sample grown in the energetic irradiation process.[15] Low temperature annealing, adopted in this report, at 525 $^0$C may not have removed the defects completely. The Raman peaks (Fig. 2b) get prominence for the annealed sample implanted at a fluence of $4 \times 10^{17}$ cm$^{-2}$. It suggests good crystalline quality of the InN phase with $E_1$(TO) peak $\sim$ 475 cm$^{-1}$ closing towards the theoretical value of 470 cm$^{-1}$ [ref. 14].

Temperature-dependent PL studies show (Fig. 3) a broad (FWHM $\sim$0.25 eV) peak around $\sim$1.06 eV at temperatures $\leq$ 150K and designated as direct band-to-band transition peak corresponding to InN phase.[1,16] A broad PL peak in InN phase is nothing unusual, as also reported in very good quality InN thin film.[17] The expected blue shift of the band-to-band transition peak with decreasing temperature is not observed in case of InN. This may be due to the strain, which compensate for the expected shift in band-to-band peak position in the embedded InN structures. On the other hand, the PL peak (Fig. 3) around $\sim$1.34 eV at 300K corresponds to direct band gap of InP phase and shows an expected blue shift with decreasing temperature. The confirmation for direct band-to-band transition peak around $\sim$1.06 eV comes from our power-dependent PL study (Fig. 4). The power-dependent linear variation of PL intensities around 1.06 eV (inset Fig. 4), proves that it is indeed a direct band-to-band transition peak for grown InN phase. This peak position around 1.06 eV at 15K is some what higher value than that reported (~0.66-0.74 eV) by J. Wu *et al.*[17] Increase in band gap value may be correlated to the increase in *n*-type carrier concentration (Burstein-Moss effect)[18] introduced by nitrogen vacancy in the grown InN phase.[19] A short annealing time in the embedded structure bound to have nitrogen deficiency. A short annealing time is preferred to avoid major depletion of P in InP lattice and to get rid of complexities in structures of grown InN phase.

In conclusion, high-quality InN phase is grown by a two-step technique comprising of low N$^+$ beam current implantation in InP(100) substrates at an elevated temperature and subsequent low temperature annealing treatment in N$_2$ ambient. InN phase is grown for the fluence above $2 \times 10^{17}$ cm$^{-2}$. Raman studies show good crystalline quality for the sample grown at a fluence of $4 \times 10^{17}$ cm$^{-2}$. Temperature- and power-dependent photoluminescence studies show direct band-to-band transition peak around ~1.06 eV corresponding to InN phase.

We acknowledge J. C. George, MSD for the design and fabrication of the heater arrangement used in the high temperature irradiation. We also thank C. S. Sundar, MSD for his encouragement in pursuing this work.

**References :**

1. J. Wu, W. Walukiewicz, K. M. Yu, J. W. Ager III, E. E. Haller, H. Lu and W. J. Schaff, Y. Saito and Y. Nanishi, Appl. Phys. Lett. **80**, 3967 (2002).

2. S. K. O'Leary B. E. Foutz, M. S. Shur, U. V. Bhapkar, and L. F. Eastman, J. Appl. Phys. **83**, 826 (1998).

3. E. N. Matthias, and B. M. Allen, IEEE Trans. Electron Devices **ED-34**, 257 (1987).

4. M. E. Jones J. R. Shealy, and J. R. Engstrom, Appl. Phys. Lett. **67**, 542 (1995).

5. T. Makimoto T. Makimoto, and N. Kobayashi Appl. Phys. Lett **70**, 1161 (1997).

6. X. W. Lin, M. Behar, and R. Maltez, W. Swider, Z. Liliental-Weber, and J. Washburn, Appl. Phys. Lett. **67**, 2699 (1995).

7. S. Dhara, P. Magudapathy, R. Kesavamoorthy, S. Kalavathi, K. G. M. Nair, G. M. Hsu, L. C. Chen, K. H. Chen, K. Santhakumar and T. Soga, Appl. Phys. Lett. **87**, 261915 (2005).

8. K. Kuriyama, H. Kondo, N. Hayashi, M. Ogura, M. Hasegawa, N. Kobayashi, Y. Takahashi, and S. Watanabe, Appl. Phys. Lett. **79**, 2546 (2001).

9. Y. Suzuki, H. Kumano, W. Tomota, N. Sanada, and Y. Fukuda Appl. Surf. Sci **162-163**, 172 (2000).

10. J. P. Pan, A. T. S. Weey, C. H. A. Huan, H. S. Tan, and K. L. Tan, J. Phys D: Appl. Phys. **29**, 2997 (1996).

11. K. Santhakumar R. Kesavamoorthy, K.G.M. Nair, P. Jayavel, D. Kanjilal, V. S. Sastry, and V. Ravichandran, Nucl. Instru. & Meth. in Phys. Res. B **212**, 521 (2003).

12. J. F. Ziegler, J. P. Biersack, and U. Littmark, *The Stopping and Range of Ions in Solids* (Pergamon, New York, 1985); http://www.srim.org/

13. JCPDS card numbers corresponding to phases; InP (**32-0452**) and InN (**02-1450**).

14. G. Kaczmarczyk, A. Kaschner, S. Reich, A. Hoffmann, C. Thomsen, D. J. As, A. P. Lima, D. Schikora, K. Lischka, R. Averbeck, and H. Riechert,, Appl. Phys. Lett. **76**, 2122 (2000).

15. Z. G. Qian, W. Z. Shen, H. Ogawa, and Q. X. Guo, J. Appl. Phys. **93**, 2643 (2003).

16. T. Matsuok, H. Okamoto, M. Nakao, H. Harima, and E. Kurimoto, Appl. Phys. Lett. **81**, 1246 (2002).

17. J. Wu, W. Walukiewicz, W. Shan, K. M. Yu, J. W. Ager III, S. X. Li, E. E. Haller, H. Lu and J. Schaff, J. Appl. Phys. **94**, 4457 (2003).

18. E. Burstein, Phys. Rev. **93**, 632 (1954).

19. H. P. Maruska, and J. J. Tietjen, Appl. Phys. Lett. **78**, 3008 (1995).

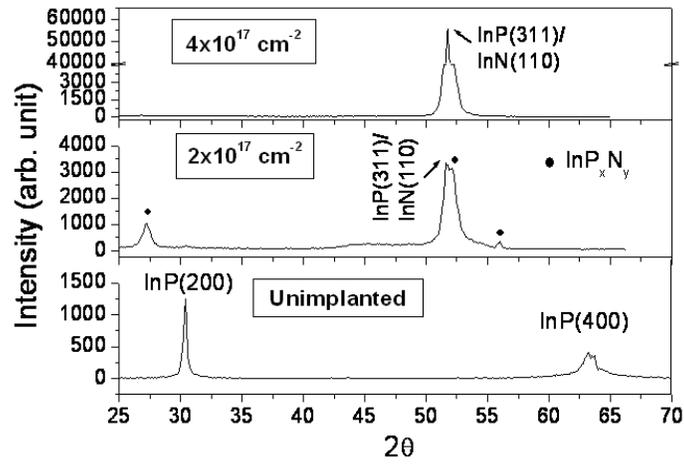

Fig. 1. GIXRD study of the unimplanted, and annealed samples irradiated at fluences of $2\times10^{17}$ cm$^{-2}$ and $4\times10^{17}$ cm$^{-2}$.

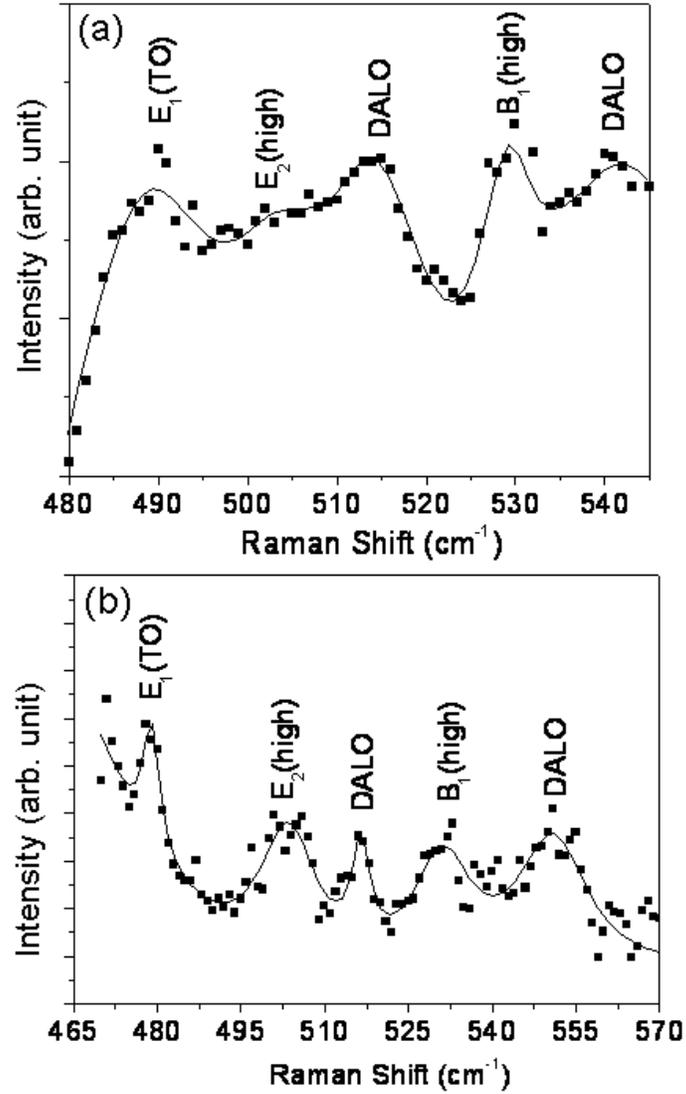

Fig. 2. Raman scattering study of annealed samples irradiated at fluences of a) $2 \times 10^{17}$ cm$^{-2}$ and b) $4 \times 10^{17}$ cm$^{-2}$. Points are the data and the solid line is the fit to Lorentzian line shape function.

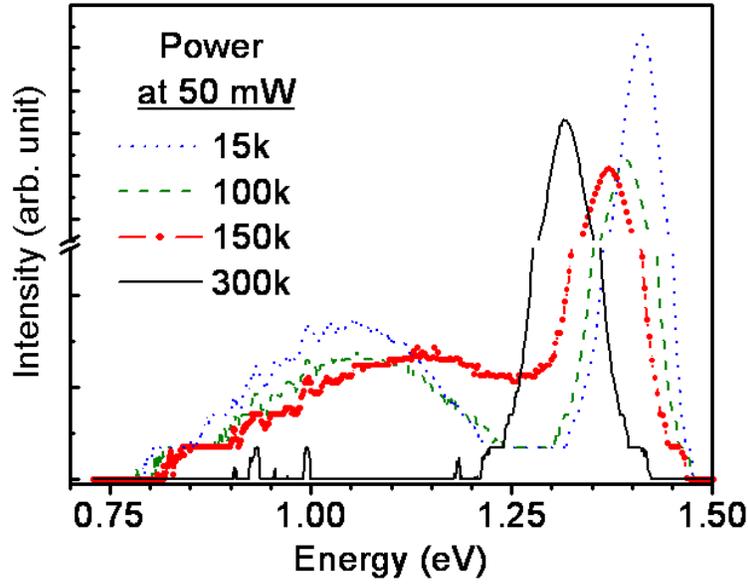

Fig. 3. Temperature-dependent PL spectra for the post-annealed InN implanted with a fluence of $4 \times 10^{17}$ cm$^{-2}$ showing presence of band-to band transition peak ~1.06 eV.

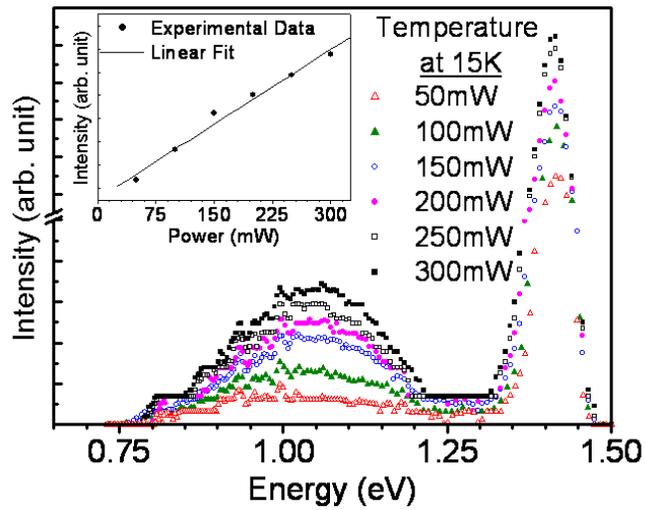

Fig. 4. Excitation power-dependent PL spectra for the post-annealed InN implanted with a fluence of $4 \times 10^{17}$ cm$^{-2}$. Inset shows variation of PL peak (at ~1.06 eV) intensities with the excitation power and the linear fitting.